\documentstyle[12pt]{article}

\input epsf.sty

\def\cm{{\cal M}}

\def\pp{{\mathchoice
          {
              \kern 1pt%
              \raise 1pt
              \vbox{\hrule width5pt height0.4pt depth0pt
                    \kern -2pt
                    \hbox{\kern 2.3pt
                          \vrule width0.4pt height6pt depth0pt
                          }
                    \kern -2pt
                    \hrule width5pt height0.4pt depth0pt}%
                    \kern 1pt
           }
            {
              \kern 1pt%
              \raise 1pt
              \vbox{\hrule width4.3pt height0.4pt depth0pt
                    \kern -1.8pt
                    \hbox{\kern 1.95pt
                          \vrule width0.4pt height5.4pt depth0pt
                          }
                    \kern -1.8pt
                    \hrule width4.3pt height0.4pt depth0pt}%
                    \kern 1pt
            }
            {
              \kern 0.5pt%
              \raise 1pt
              \vbox{\hrule width4.0pt height0.3pt depth0pt
                    \kern -1.9pt  
                    \hbox{\kern 1.85pt
                          \vrule width0.3pt height5.7pt depth0pt
                          }
                    \kern -1.9pt
                    \hrule width4.0pt height0.3pt depth0pt}%
                    \kern 0.5pt
            }
            {
              \kern 0.5pt%
              \raise 1pt
              \vbox{\hrule width3.6pt height0.3pt depth0pt
                    \kern -1.5pt
                    \hbox{\kern 1.65pt
                          \vrule width0.3pt height4.5pt depth0pt
                          }
                    \kern -1.5pt
                    \hrule width3.6pt height0.3pt depth0pt}%
                    \kern 0.5pt
            }
        }}

\def\mm{{\mathchoice
                       {
                             \kern 1pt
               \raise 1pt    \vbox{\hrule width5pt height0.4pt depth0pt
                                  \kern 2pt
                                  \hrule width5pt height0.4pt depth0pt}
                             \kern 1pt}
                       {
                            \kern 1pt
               \raise 1pt \vbox{\hrule width4.3pt height0.4pt depth0pt
                                  \kern 1.8pt
                                  \hrule width4.3pt height0.4pt depth0pt}
                             \kern 1pt}
                       {
                            \kern 0.5pt
               \raise 1pt
                            \vbox{\hrule width4.0pt height0.3pt depth0pt
                                  \kern 1.9pt
                                  \hrule width4.0pt height0.3pt depth0pt}
                            \kern 1pt}
                       {
                           \kern 0.5pt
             \raise 1pt  \vbox{\hrule width3.6pt height0.3pt depth0pt
                                  \kern 1.5pt
                                  \hrule width3.6pt height0.3pt depth0pt}
                           \kern 0.5pt}
                       }}

\def\ad{{\kern0.5pt
                   \alpha \kern-5.05pt \raise5.8pt\hbox{$\textstyle.$}\kern
0.5pt}}

\def\bd{{\kern0.5pt
                   \beta \kern-5.05pt \raise5.8pt\hbox{$\textstyle.$}\kern
0.5pt}}

\def\qd{{\kern0.5pt
                   q \kern-5.05pt \raise5.8pt\hbox{$\textstyle.$}\kern
0.5pt}}
\def\Dot#1{{\kern0.5pt
     {#1} \kern-5.05pt \raise5.8pt\hbox{$\textstyle.$}\kern
0.5pt}}

\def\fracm#1#2{\hbox{\large{${\frac{{#1}}{{#2}}}$}}}

\skewchar\fivmi='177
\skewchar\sixmi='177
\skewchar\sevmi='177
\skewchar\egtmi='177
\skewchar\ninmi='177
\skewchar\tenmi='177
\skewchar\elvmi='177
\skewchar\twlmi='177
\skewchar\frtnmi='177
\skewchar\svtnmi='177
\skewchar\twtymi='177
\def\@magscale#1{ scaled \magstep #1}
\skewchar\fivsy='60
\skewchar\sixsy='60
\skewchar\sevsy='60
\skewchar\egtsy='60
\skewchar\ninsy='60
\skewchar\tensy='60
\skewchar\elvsy='60
\skewchar\twlsy='60
\skewchar\frtnsy='60
\skewchar\svtnsy='60
\skewchar\twtysy='60

\catcode`@=11
\def\un#1{\relax\ifmmode\@@underline#1\else
        $\@@underline{\hbox{#1}}$\relax\fi}
\catcode`@=12

\def\a{\alpha}
\def\b{\beta}

\def\d{\delta}
\def\e{\epsilon}

\def\g{\gamma}

\def\l{\lambda}
\def\m{\mu}

\def\o{\omega}

\def\q{\theta}

\def\s{\sigma}

\def\z{\zeta}
\def\D{\Delta}

\def\S{\Sigma}

\def\cm{{\cal M}}

\def\dslash{\not{\hbox{\kern-2pt $\partial$}}}
\def\Dslash{\not{\hbox{\kern-4pt $D$}}}
\def\pslash{\not{\hbox{\kern-2.3pt $p$}}}
 \newtoks\slashfraction
 \slashfraction={.13}
 \def\slash#1{\setbox0\hbox{$ #1 $}
 \setbox0\hbox to \the\slashfraction\wd0{\hss \box0}/\box0 }
\def\slsh{\!\bigm|} 
 
\font\ro=cmsy10                          
\def\kcr{{\hbox{\ro \char'170}}}                
\def\ktl{{\hbox{\ro \char'170}}}        
\def\ktr{{\hbox{\ro \char'170}}}        
\def\kbl{{\hbox{\ro \char'170}}}        
\def\kbr{{\hbox{\ro \char'170}}}        

\def\plpl{\raise-2pt\hbox{$\raise3pt\hbox{$_+$}\hskip-6.67pt\raise0.0pt
\hbox{$^+$}\hskip 0.01pt$}}
\def\mimi{\raise-2pt\hbox{$\raise3pt\hbox{$_-$}\hskip-6.67pt\raise0.0pt
\hbox{$^-$}\hskip 0.01pt$}} 

\def\bo{{\raise.15ex\hbox{\large$\Box$}}}               
\def\pa{\partial}                                       
\def\de{\nabla}                                         
\def\TH{{\raise.2ex\hbox{$\displaystyle \bigodot$}\mskip-4.7mu \llap H \;}}
\def\face{{\raise.2ex\hbox{$\displaystyle \bigodot$}\mskip-2.2mu \llap {$\ddot
        \smile$}}}                                      

\def\sp#1{{}^{#1}}                              
\def\Tilde#1{\widetilde{#1}}                    
\def\Hat#1{\widehat{#1}}                        
\def\Bar#1{\overline{#1}}                       
\def\leftrightarrowfill{$\mathsurround=0pt \mathord\leftarrow \mkern-6mu
        \cleaders\hbox{$\mkern-2mu \mathord- \mkern-2mu$}\hfill
        \mkern-6mu \mathord\rightarrow$}
\def\dvec#1{\vbox{\ialign{##\crcr
        \leftrightarrowfill\crcr\noalign{\kern-1pt\nointerlineskip}
        $\hfil\displaystyle{#1}\hfil$\crcr}}}           

\def\fracm#1#2{\hbox{\large{${\frac{{#1}}{{#2}}}$}}}
\def\frac#1#2{{\textstyle{#1\over\vphantom2\smash{\raise.20ex
        \hbox{$\scriptstyle{#2}$}}}}}                   
\def\ha{\frac12}                                        
\def\sfrac#1#2{{\vphantom1\smash{\lower.5ex\hbox{\small$#1$}}\over
        \vphantom1\smash{\raise.4ex\hbox{\small$#2$}}}} 
\def\bfrac#1#2{{\vphantom1\smash{\lower.5ex\hbox{$#1$}}\over
        \vphantom1\smash{\raise.3ex\hbox{$#2$}}}}       
\def\afrac#1#2{{\vphantom1\smash{\lower.5ex\hbox{$#1$}}\over#2}}    

\newskip\humongous \humongous=0pt plus 1000pt minus 1000pt
\def\caja{\mathsurround=0pt}
\def\eqalign#1{\,\vcenter{\openup2\jot \caja
        \ialign{\strut \hfil$\displaystyle{##}$&$
        \displaystyle{{}##}$\hfil\crcr#1\crcr}}\,}
\newif\ifdtup

\def\ref#1{$\sp{#1)}$}

\parskip=\medskipamount                 
\thicklines                         

\def\oldheadpic{                                
        \setlength{\unitlength}{.4mm}
        \thinlines
        \par
        \begin{picture}(349,16)
        \put(325,16){\line(1,0){4}}
        \put(330,16){\line(1,0){4}}
        \put(340,16){\line(1,0){4}}
        \put(335,0){\line(1,0){4}}
        \put(340,0){\line(1,0){4}}
        \put(345,0){\line(1,0){4}}
        \put(329,0){\line(0,1){16}}
        \put(330,0){\line(0,1){16}}
        \put(339,0){\line(0,1){16}}
        \put(340,0){\line(0,1){16}}
        \put(344,0){\line(0,1){16}}
        \put(345,0){\line(0,1){16}}
        \put(329,16){\oval(8,32)[bl]}
        \put(330,16){\oval(8,32)[br]}
        \put(339,0){\oval(8,32)[tl]}
        \put(345,0){\oval(8,32)[tr]}
        \end{picture}
        \par
        \thicklines
        \vskip.2in}
\def\oldtitle#1#2#3#4{\oldheadpic\begin{center}\vglue.5in{\large\bf #1}\\[.6in]
        {#2}\\[.1in] {\it Department of Physics and Astronomy}\\
        {\it University of Maryland, College Park, MD 20742}\\[.6in]
        Physics Publication \#{#3}\\ {#4}\\[1.5in] {\bf ABSTRACT}\\[.1in]
        \end{center} \begin{quotation}}                 
\def\oldTitle#1#2#3#4#5#6#7{\oldheadpic\begin{center} \vglue .4in
        {\large\bf #1}\\[.4in]
        {#2}\\[.1in] {\it Department of Physics and Astronomy}\\
        {\it University of Maryland, College Park, MD 20742}\\[.1in]
        {#3}\\[.1in] {\it {#4}}\\ {\it {#5}}\\[.4in]
        Physics Publication \#{#6}\\ {#7}\\[.5in] {\bf ABSTRACT}\\[.1in]
        \end{center} \begin{quotation}}                 
\def\border{                                            
        \setlength{\unitlength}{1mm}
        \newcount\xco
        \newcount\yco
        \xco=-21
        \yco=12
        \begin{picture}(140,0)
        \put(\xco,\yco){$\ktl$}
        \advance\yco by-1
        {\loop
        \put(\xco,\yco){$\kcr$}
        \advance\yco by-2
        \ifnum\yco>-240
        \repeat
        \put(\xco,\yco){$\kbl$}}
        \xco=158
        \yco=12
        \put(\xco,\yco){$\ktr$}
        \advance\yco by-1
        {\loop
        \put(\xco,\yco){$\kcr$}
        \advance\yco by-2
        \ifnum\yco>-240
        \repeat
        \put(\xco,\yco){$\kbr$}}
        \put(-20,13){\tiny University of Maryland Elementary Particle
Physics University of Maryland Elementary Particle Physics University of
Maryland Elementary Particle Physics}
        \put(-20,-241.5){\tiny University of Maryland Elementary
Particle Physics University of Maryland Elementary Particle Physics
University of Maryland Elementary Particle Physics}
        \end{picture}
        \par\vskip-8mm}
\def\bordero{                                           
        \setlength{\unitlength}{1mm}
        \newcount\xco
        \newcount\yco
        \xco=-31
        \yco=12
        \begin{picture}(140,0)
        \put(\xco,\yco){$\ktl$}
        \advance\yco by-1
        {\loop
        \put(\xco,\yco){$\kclr}
        \advance\yco by-2
        \ifnum\yco>-240
        \repeat
        \put(\xco,\yco){$\kbl$}}
        \xco=151
        \yco=12
        \put(\xco,\yco){$\ktr$}
        \advance\yco by-1
        {\loop
        \put(\xco,\yco){$\kcr$}
        \advance\yco by-2
        \ifnum\yco>-240
        \repeat
        \put(\xco,\yco){$\kbr$}}
        \put(-20,12){\ooo bacdefghidfghghdhededbihdgdfdfhhdheidhdhebaaahjhhdahba

hgdedge
   hgfdiehhgdigicba}
        \put(-20,-241.5){\ooo ababaighefdbfghgeahgdfgafagihdidihiidhiagfedhadbfd

ecdcdfa
   gdcbhaddhbgfchbgfdacfediacbabab}
        \end{picture}
        \par\vskip-8mm}
\def\headpic{                                           
        \indent
        \setlength{\unitlength}{.4mm}
        \thinlines
        \par
        \begin{picture}(29,16)
        \put(165,16){\line(1,0){4}}
        \put(170,16){\line(1,0){4}}
        \put(180,16){\line(1,0){4}}
        \put(175,0){\line(1,0){4}}
        \put(180,0){\line(1,0){4}}
        \put(185,0){\line(1,0){4}}
        \put(169,0){\line(0,1){16}}
        \put(170,0){\line(0,1){16}}
        \put(179,0){\line(0,1){16}}
        \put(180,0){\line(0,1){16}}
        \put(184,0){\line(0,1){16}}
        \put(185,0){\line(0,1){16}}
        \put(169,16){\oval(8,32)[bl]}
        \put(170,16){\oval(8,32)[br]}
        \put(179,0){\oval(8,32)[tl]}
        \put(185,0){\oval(8,32)[tr]}
        \end{picture}
        \par\vskip-6.5mm
        \thicklines}
\def\title#1#2#3#4{\border\headpic {\hbox to\hsize{#4 \hfill UMDEPP #3}}\par
        \begin{center} \vglue .5in {\large\bf #1}\\[.6in]
        {#2}\\[.1in] {\it Department of Physics and Astronomy}\\
        {\it University of Maryland, College Park, MD 20742}\\[1.5in]
        {\bf ABSTRACT}\\[.1in] \end{center} \begin{quotation}}  
\def\Title#1#2#3#4#5#6#7{\border\headpic
        {\hbox to\hsize{#7 \hfill UMDEPP #6}}\par
        \begin{center} \vglue .4in {\large\bf #1}\\[.4in]
        {#2}\\[.1in] {\it Department of Physics and Astronomy}\\
        {\it University of Maryland, College Park, MD 20742}\\[.1in]
        {#3}\\[.1in] {\it {#4}}\\ {\it {#5}}\\[.5in] {\bf ABSTRACT}\\[.1in]
        \end{center} \begin{quotation}}                 
\def\endtitle{\end{quotation}\newpage}                  

\begin{document}

\border\headpic {\hbox to\hsize{September 1997 \hfill UMDEPP 98-13}}\par
\begin{center}
\vglue .4in
{\large\bf Ectoplasm Has No Topology: The Prelude\footnote{
Research supported by NSF grant \# PHY-96-43219}
${}^,$ \footnote {Supported in part by NATO Grant CRG-93-0789} 
}\\[.4in]
S. James Gates, Jr.\footnote{gates@umdhep.umd.edu} \\[.1in]
{\it Department of Physics\\
University of Maryland at College Park\\
College Park, MD 20742-4111, USA} 
\\[0.55in]
A Presentation at the\\
International Seminar on\\
``Supersymmetries and Quantum Symmetries''\\
in memory of\\
Victor I. Qgievetsky\\
Dubna, Russia\\
July 22 - 26, 1997 
\\[1.0in]

{\bf ABSTRACT}\\[.1in]
\end{center}
\begin{quotation}

Preliminary evidence is presented that a long overlooked and critical
element in the fundamental definition of a general theory of integration 
over curved Wess-Zumino superspace lies with the imposition of
``the Ethereal Conjecture'' which states the necessity of the superspace
to be topologically ``close'' to its purely bosonic sub-manifold.  
As a step in proving this, a new theory of integration of closed super 
p-forms is proposed.

\endtitle

\pagebreak

Presently `Salam-Strathdee superspace \cite{salst}' is almost universally 
accepted as the requisite mathematical setting for describing supersymmetrical 
field theories.  Even so, there remain a fairly large number of open 
questions about superspace, particularly with regard to those with large 
values of N ($\equiv N_F$) or D ($\equiv N_B$).  There are also 
particularly pointed questions that remain largely unanswer about a 
general theory of integration on the curved versions of these spaces 
also known as `Wess-Zumino superspace \cite{wz}.'  To answer some of 
these questions, extensions such as `harmonic superspace' have been 
developed especially by the late Dr.~Ogievetsky and collaborators. 
Although my discussion today will only tangentially touch on such 
constructions, I wish to dedicate this talk to Victor Isaakovich's 
memory.

Near the beginning of research using superspace, more mathematically 
motivated investigators such as Rogers \cite{rog} asked a question we
may paraphrase as,

${~~~~~~~~}$``Is it possible to construct a superspace whose topological 
properties \newline \indent ${~~~~~~~~~~}$are significantly different from 
those of its purely bosonic subspace?'' 

\noindent In all cases of interest to physicists to date the answer 
appears to be, ``No!"  The emphasis on this negation is mine own because
I believe that there is a hidden message in this answer.

In establishing a nomenclature appropriate to researching these issues, one
often finds the `spiritualist' denotations (see for example \cite{dwt})

${~~~~~~~~~~~~~~~~}$ monomials in $x$ $\equiv$ body of the superspace,

${~~~~~~~~~~~~~~~~}$ monomials in $x$ and $\q$ or purely 
$\q$ $\equiv$ soul of the superspace.

\noindent In deference to this convention, I may call the `basic substance'
of which the soul is composed, the ``ectoplasm'' of superspace.  

There is a peculiar sense in which the question of how to construct 
integration measures over \underline{curved} superspaces is unanswered.   
Arnowitt, Nath and Zumino \cite{anz} first suggested such integration 
measures should be written as

$$\int~d\mu ~\equiv~ \int d^{N_B + N_F} z ~ {\rm E}^{-1} ~=~
\int d^{N_B + N_F} z ~ [ sdet \, ({\rm E}_{\un A} {}^{\un M} 
 (\q , x)\,) ]^{-1} ~~~.
\eqno(1) 
$$
for a superspace of $N_{B}$ bosonic coordinate and $N_F$ fermionic
coordinates. 

In principle this is perfectly consistent.  In practice, however, for any
theory with large $N_{B}$ or $N_{F}$ ($N_{F}=4$ is large), this becomes an
impractical way to obtain component results in a supergravity theory of
`physical' interest.  The impracticality arises because the complete 
$\q$-expansion of the superdeterminant of the inverse vielbein $[ 
sdet \, ({\rm E}_{\un A} {}^{\un M}  (\q , x)\,) ]^{-1}$ is 
complicated to calculate\footnote{To my knowledge, this calculation has 
only been done explicitly by no more than six physicists \newline 
${~\,~~~}$ to this date for 4D, N = 1 supergravity.}.  For practical 
calculations an alternative to the method of Arnowitt, Nath and Zumino 
is required.  To my knowledge, only two such alternatives exist 
in the literature.  They have been discussed in three books 
listed by authors below.

\begin{description}
\item{${~~~~~}$ a.}  ``Covariant Theta Expansion'' - Wess 
\& Bagger, \cite{bw}

\item{${~~~~~}$ b.}  ``Density Projectors" - Gates, Grisaru, Ro\v{c}ek 
\& Siegel, \cite{ggrs}
\newline ${~~~~~~~~~~~~~~~~~~~~~~~~~~~~~~~\,}$ - Buchbinder \& Kuzenko.
\cite{bk}
\end{description}

\noindent I will obviously speak on the second of these because I 
have recently found increasing and unexpected indications that it is 
directly connected to more general issues of the calculus and topology 
of curved supermanifolds with torsion.

I begin by writing the ``Ectoplasmic Integration Theorem'' (or  E.I.T.).  
There should exist an operator ${\cal D}^{N_F}$ such that
$$ 
\int d^{N_B + N_F} z ~ {\rm E}^{-1} {\cal L} ~=~ \int d^{N_B} z ~
{\rm e}^{-1} \, [~ {\cal D}^{N_F} {\cal L} \slsh ~] ~~~,
\eqno(2)
$$
independent of the superfield $\cal{L}$ that appears in this equation
and where
$$
{\rm e}^{-1} ~ \equiv ~ [det\, ({\rm e}_{\un a} {}^{\un m} 
( x)\,)] ^{-1}  ~~~,~~~ {\cal D}^{N_F} {\cal L} | ~ \equiv ~ 
\lim_{\q \to 0}^{} ~ (\, {\cal D}^{N_F} {\cal L} 
) ~~~. \eqno(3)
$$
\noindent
This theorem is of a similar form to that of the standard Gauss', Green's
or Stoke's Theorems of multi-variable calculus.  It is different, however, 
because the operator ${\cal D}^{N_F}$ appears on the ``wrong'' side of 
the equation from the standard multi-variable calculus analogs. The
E.I.T. is also the natural extension of the Berezinian definition
of integrating over Grassmann numbers \cite{bz}.

To see why this is a practical improvement in calculational matters, let 
me consider the case of flat 4D, N = 1 superspace where the E.I.T. 
becomes
$$
\int d^{4} x \, d^2 \q \, d^2 {\bar \q} ~ {\cal L} ~\equiv~ \fracm 12 \,
\Big\{ \int d^{4} x ~ [\, D^2 \, {\Bar D}{}^2  \,  {\cal L} \, | ~] 
~+~{\rm {h. \, c.}} ~ \Big\} ~~~,
\eqno(4) 
$$
where
$$
D_{\a}~\equiv~ \partial_{\alpha} ~ +~ i \frac{1}{2} \bar{\q}^{
\Dot{\alpha}} \partial_{\underline{a}}  ~~~,~~~{ \Bar D}_{\Dot{\alpha}} 
~\equiv~ \bar{\partial}_{\Dot{\alpha}}~+~i \frac{1}{2} \q^{\alpha} 
\partial_{\underline{a}}  ~~~.
\eqno(5) 
$$
Anyone familiar with rigid supersymmetry can attest to the practical 
utility of the above equation.  For example, if I define ${\cal L} ~ 
\equiv~ {\Bar \Phi} \Phi $ where $\Bar{D}_{\Dot{\alpha}} \, \Phi~=~0$
use the component field definitions $A(x) \equiv \Phi |$, $\psi_{
\alpha} (x) \equiv D_{\alpha} \Phi |$ and ${F(x)} \equiv D^2 \Phi |$, 
apply the E.I.T. and use of the Leibnitz rule for
differentiation, it is simple to show
$$
\int d^{4} x \, d^2 \q \, d^2 {\bar \q} ~ \Bar{\Phi} \Phi~=~ \int d^4~ x
~[~ - \frac{1}{2} (\partial^{\underline{a}} \Bar{A} )\, (\partial_{
\underline{a}} A)~-~i \Bar{\psi}{}^{\Dot{\alpha}} \, \partial_{\underline{a}} 
\psi^{\alpha}~+~ F \Bar{F} ~] ~~~. \eqno(6)
$$
{\it {No explicit}} $\q$-{\it {expansion was required at any point to 
derive this component result}}.  Thus, it should be obvious why it is 
calculationally superior to use the E.I.T.  By using techniques that 
are essentially the same as above, we simple by-pass the need to know the 
explicit structure of the $\q$-expansion of $[ sdet \, ({\rm E}_{\un A} 
{}^{\un M}  (\q , x)\,) ]^{-1}$!

From this viewpoint, the whole problem becomes how to develop a theory 
for the calculation of the operator ${\cal D}^{N_{F}}$ that appears in 
equation (2).  The expression $ e^{-1} [{\cal D}^{N_{F}} \cal{L} |
\, ] $ is called ``the density projection operator'' or ``density 
projector'' (see `{\it {Superspace}}' \cite{ggrs1} or `{\it {Ideas}}'
\cite{bk1} ).  It should be clear that this 
operator, in the general case, can be written as
$$
\int d^{N_B} z ~{\rm e}^{-1} \,  \Big[ ~ {\cal D}^{N_F} {\cal 
L} \, | ~ \Big] ~=~ \int d^{N_B} z ~{\rm e}^{-1} \, \Big[ ~ 
\sum_{i = 0}^{N_F} \, c_{(N_F - i)} \, (\nabla \, \cdot\,  \, 
\cdot\, \, \cdot\, \nabla)^{N_F - i}  \, {\cal L} | ~ \Big]
~~~,
\eqno(7) $$
in terms of some field-dependent coefficients $c_{(N_{F}-i)}$
and powers of the spinorial superspace supergravity covariant 
derivative $\nabla_{\a}$.  How are these coefficients to be found?  

In `{\it {Superspace}}' \cite{ggrs1} it was shown that given the 
local supersymmetry variations of some matter superfield, it is possible 
to re-construct these coefficients.  In `{\it {Ideas}}' \cite{bk1},
it was shown that the density projector follows after solving
the constraints to find the basic supergravity pre-potentials.
Neither of these approaches is a theory\footnote{We may think of 
the approach in \cite{ggrs1} as a `handicraft' method for summarizing 
com- \newline ${~~~~\,}$ ponent results. The fact that it was required 
to go component at all, was equivalent \newline ${~~~~\,}$ 
to an admission that we did not have an {\it a} {\it {priori}} 
theoretical basis for this result.} for ${\cal D}^{N_F}$.  In the 
early to middle eighties, Zumino was the first to raise the question 
of a purely {\it {theoretical}} basis for this operator. This bring us 
to the point of my presentation.

In the rest of my presentation, I will attempt to convince the
reader that the answer can be found in the study of super topology 
similar to the investigations by Rogers.  I will argue that local 
supergravity theories (as a principle) obey what I call ``The Ethereal 
Conjecture'' which largely determines the form of ${\cal D}^{N_F}$.

\begin{figure}[htbp]
\begin{center}
\centerline{\epsfbox{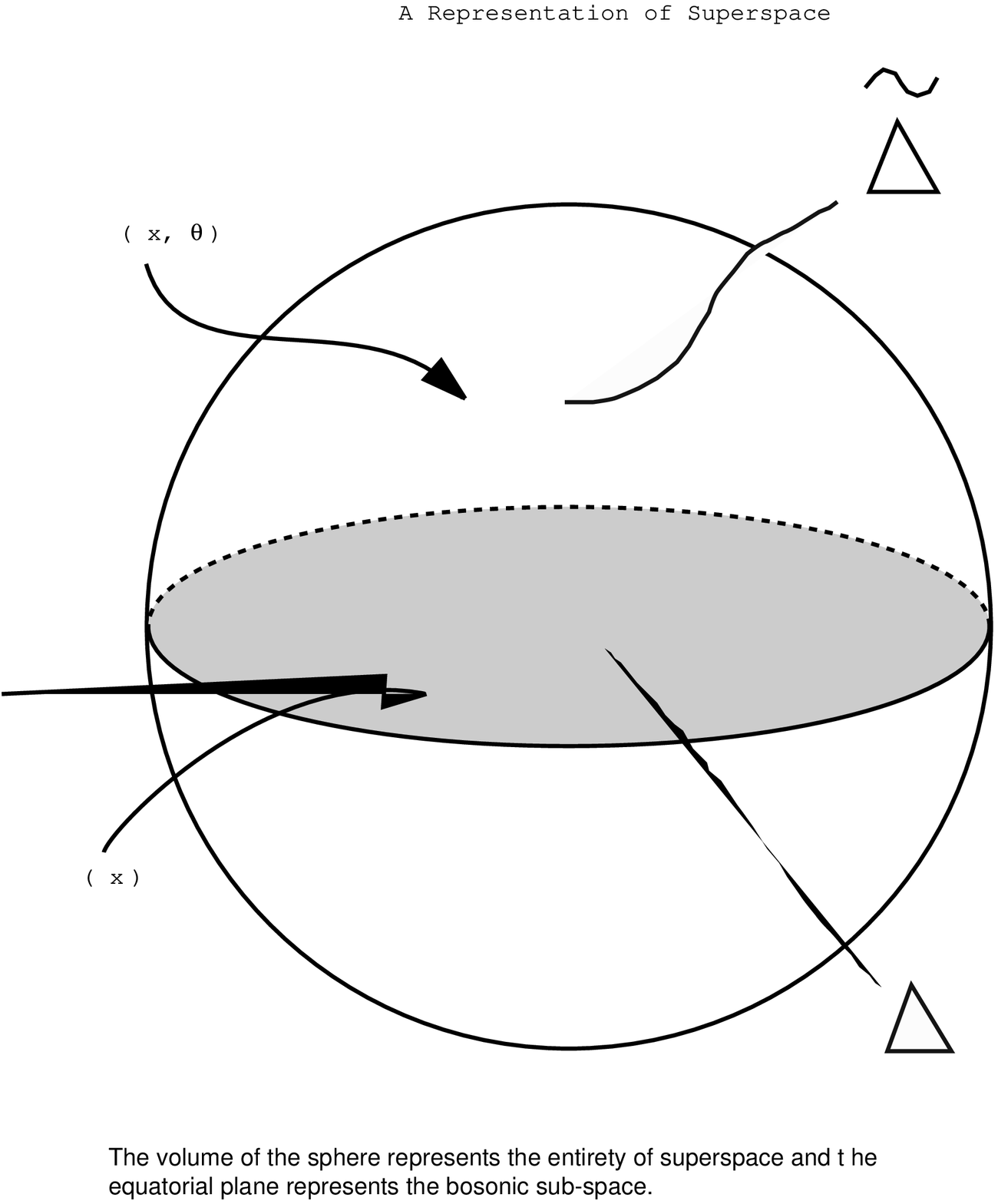}}
${~}$
\end{center}
\end{figure}

\newpage
In terms of the topological indices represented in the
diagram I may formulate the Ethereal Conjecture (E.~C.) as:

{\it {${~~~~~~}$ For all Wess-Zumino superspaces, the operator 
${\cal D}^{N_F}$ that appears \newline ${~~~~~~~}$ in the E.~I~.T.~has 
the property that it insures that}} ${\Hat \D}\ \simeq\ \D$.  

\noindent I will later give some meaning to $\simeq$ that appears
in this relation.  This is the result to which I was alluding in 
the other-worldly sounding title.  Stated another way, this
relation says that the local integration measure of a Wess-Zumino
superspace is such that the superspace is topologically ``close'' 
to its underlying bosonic manifold. Or alternately, the
local Grassmann integration measure of a Wess-Zumino superspace 
is actually to be detemined from topological considerations.

For an ordinary $N_{B}$-form $f_{{{\underline{a}}_{1}}\ \cdots\
{\underline{a}}_{{N}_{B}}}$ defined over an ordinary bosonic manifold, 
the calculation of the index $\Delta$ is just given by
$$
\D ~\equiv~ ({N_B}!)^{-1} \, 
\int d^{N_B} z ~{\rm e}^{-1} \, \e^{{\un a}_1 \cdots {\un a}_{N_B} }
f_{{\un a}_1 \cdots {\un a}_{N_B} } ~~~,
\eqno(8)
$$
and for $\D$ to truly correspond to a topological index, we must have
$f_{{{\underline{a}}_{1}}\ \cdots\ {\underline{a}}_{{N}_{B}}}$ satisfy a 
Bianchi-type identity (for the purely bosonic case this is trivial)
$$
{\rm e}_{[ {\un a}_1 | } f_{ | {\un a}_2 \cdots {\un a}_{1 + N_B} ]} 
~-~ c_{ [ {\un a}_1 \, {\un a}_2 | } {}^{\un d} \,f_{ 
{\un d}\,  | {\un a}_3 \cdots {\un a}_{1 + N_B } ] } ~=~ 0 ~~~,
\eqno(9)
$$
and as well $f_{ {\un a}_1 \ \cdots\ {\un a}_{N_B} }$ must 
\underline{not} be globally defined over the entire $N_B$-dimensional 
bosonic manifold. In this equation ${\rm e}_{ {\un a}  }$ denotes
the local frame field operator of the bosonic sub-manifold and $c_{ 
{\un a} \, {\un b} } {}^{\un c}$ denotes the associated anholonomy.

What are the corresponding structures available over a Salam-Strathdee
superspace?

In 1981 \cite{sjg1}, I proposed the initial formulation of 
\underline{irreducible} off-shell super p-forms for Salam-Strathdee 
superspace as a generalization of supersymmetric Yang-Mills theory.  
Although explicitly presented in a 4D, N = 1 superspace, the general 
structure is ubiquitous to all superspaces. The distinctive feature 
of the 1981 proposal was that it showed the constraints, Bianchi 
identities, and pre-potential solutions all exist for a simplex of 
super p-forms in exactly the same way as in supersymmetric Yang-Mills 
theory.

In 1983 \cite{sjg2}, I was able to derive a further interesting result. 
If a super $N_{B}$-form like $F_{ {\un A}_1 \ \cdots\ {\un A}_{N_B} }$
satisfies a set of Bianchi identities (i.e. it is super-closed), it follows 
that \underline{independent} of its constraints and in the presence of 
supergravity, there exist a W-Z gauge where
$$
\eqalign{
\Big( \, F_{ {\un a}_1 \cdots {\un a}_{N_B} } | \, \Big) ~=~ \Big[~
&{\Tilde f}_{{\un a}_1 \cdots {\un a}_{N_B} } ~+~ \l^{(N_B , 1)}
 \psi_{[ {\un a}_1 |}{}^{\a_1} \Big( \, F_{\a_1 | {\un a}_2 \cdots 
{\un a}_{N_B} ] } | \, \Big)   \cr
&~+~ \l^{(N_B , 2)} \psi_{[ {\un a}_1|}{}^{\a_1} \psi_{ |{\un a}_2|}
{}^{\a_2} \Big( \, F_{\a_1 \a_2 | {\un a}_3 \cdots {\un a}_{N_B} ] } 
| \, \Big) \cdots \cr
&~+~ \l^{(N_B , N_B)} [\, \psi_{ {\un a}_1}{}^{\a_1} \psi_{ {\un a}_2}
{}^{\a_2} \cdots \psi_{ {\un a}_{N_B}}{}^{\a_{N_B}} \,]
\Big( \, F_{\a_1 \a_2 \cdots  \a_{N_B} } | \, \Big) ~\Big]  ~~~, }
\eqno(10) $$
here ${\Tilde f}_{{\un a}_1 \cdots {\un a}_{N_B} }$ is an ordinary bosonic 
closed $N_{B}$-form, $\psi_{\un a}{}^{\a}$ denotes the component 
gravitino field and $\l^{(N_B , i)}$ are a set of constants that are 
easily derivable.  My original derivation of this was in the context of 4D, 
N = 4 supergravity but that derivation can easily be extended to all values 
of $N_B$ and $N_F$. 

Now the interesting thing about this equation is that I can isolate      
${\Tilde f}_{{\un a}_1 \cdots {\un a}_{N_B} }$ (which can differ from
$f_{{\un a}_1 \cdots {\un a}_{N_B} }$ by exact terms) to find
$$
\eqalign{ {~~~~~}
{\Tilde f}_{{\un a}_1 \cdots {\un a}_{N_B} } ~=~ \Big[~
&\Big( \, F_{ {\un a}_1 \cdots {\un a}_{N_B} } | \, \Big)  ~-~ 
\l^{(N_B , 1)} \psi_{[ {\un a}_1 |}{}^{\a_1} \Big( \, F_{\a_1 | {\un 
a}_2 \cdots {\un a}_{N_B} ] } | \, \Big)   \cr
&~-~ \l^{(N_B , 2)} \psi_{[ {\un a}_1|}{}^{\a_1} \psi_{ |{\un a}_2|}
{}^{\a_2} \Big( \, F_{\a_1 \a_2 | {\un a}_3 \cdots {\un a}_{N_B} ] } 
| \, \Big) \cdots \cr
&~-~ \l^{(N_B , N_B)} [\, \psi_{ {\un a}_1}{}^{\a_1} \psi_{ {\un a}_2}
{}^{\a_2} \cdots \psi_{ {\un a}_{N_B}}{}^{\a_{N_B}} \,]
\Big( \, F_{\a_1 \a_2 \cdots  \a_{N_B} } | \, \Big) ~\Big]  ~~~. }
\eqno(11) $$
Upon multiplying by an $\epsilon$-tensor and integrating $(1/ {N_B} !)
\int d^{N_B} z ~{\rm e}^{-1} $, I find 
$$
{\Tilde \D} ~=~ {\Hat \D} ~~~, \eqno(12) 
$$
$$
\Tilde \D ~\equiv~ ({N_B}!)^{-1} \, \int d^{N_B} z ~{\rm e}^{-1} \, 
\e^{{\un a}_1 \cdots {\un a}_{N_B} } {\Tilde f}_{{\un a}_1 \cdots 
{\un a}_{N_B} } ~~~, \eqno(13) 
$$
$$ \eqalign{
{\Hat \D} &\equiv~ \int d^{N_B} z ~{\rm e}^{-1} \,  \, \e^{{\un a}_1 
\cdots {\un a}_{N_B} } \, \Big[ ~ (N_B !)^{-1} 
\Big( \, F_{ {\un a}_1 \cdots {\un a}_{N_B} } | \, \Big) 
\,-\, \l^{(N_B , 1)} \psi_{ {\un a}_1 }{}^{\a_1} \Big( \, F_{\a_1  {\un 
a}_2 \cdots {\un a}_{N_B}  } | \, \Big) \cr
&{~~~~~~~~~~~~~~~~~~~~~~~}~-~ \l^{(N_B , 2)} \psi_{ {\un a}_1}{}^{\a_1} 
\psi_{ {\un a}_2}{}^{\a_2} \Big( \, F_{\a_1 \a_2  {\un a}_3 \cdots 
{\un a}_{N_B}  } | \, \Big) \cdots \cr
&{~~~~~~~~~~~~~~~~~~~~~~~}~-~ \l^{(N_B , N_B)} ({N_B}!)^{-1} \,[\, 
\psi_{ {\un a}_1}{}^{\a_1} \psi_{ {\un a}_2}
{}^{\a_2} \cdots \psi_{ {\un a}_{N_B}}{}^{\a_{N_B}} \,]
\Big( \, F_{\a_1 \a_2 \cdots  \a_{N_B} } | \, \Big) ~\Big]  ~~~. 
\cr {~} \cr
} \eqno(14) $$
I now define the supertopological index $\Hat {\D}$ that was 
introduced into the diagram by asserting that equation (14) {\it {is}} 
the correct definition of how to integrate the closed super 
$N_B$-form $F_{ {\un A}_1 \cdots {\un A}_{N_B} }$ over the entirety 
of the superspace\footnote{For a previous proposal to define the
integration theory of closed super p-forms, see \newline
${~~~~\,}$ the work of ref.~\cite{zup}.}!  Since ${\Tilde f}_{ 
{\un a}_{1} \ \cdots\ {\un a}_{N_B} }$ typically differs from 
$f_{{\un a}_1 \ \cdots\ {\un a}_{N_B} }$ by exact terms we have
$$ 
\Hat {\D} ~=~ \D \, + \, \cdots\ ~~~.
\eqno(15) $$
So the definition above certainly enforces the Ethereal Conjecture but how
does this solve the problem of finding ${\cal D}^{N_F}$?

The answer lies in the fact that the field strengths superfields in 
$\Hat {\Delta}$ (i.e. the $F$'s) must be chosen to be subject to the 
constaints implied by irreducibility of the super $N_{B}$-form.  In this 
case a number of the $F$'s vanish and the remaining ones, via the solution 
of their Bianchi identities, are related by $\nabla_{\a}$, the spinorial 
derivative.  When this solution for the various components of $F$ is 
inserted into ${\Hat \D}$, as if by magic the operator ${\cal D}^{N_F}$ 
appears in all the cases I have studied. Let me show by some explicit 
examples how this topological tool works.

The simplest of all 2D supergravity theories is (1,0) or heterotic 
supergravity \cite{bgm} which is described by a
set of covariant derivatives $(\nabla_{+},\ \nabla_{\mm},\ \nabla_{\pp})$
satisfying the commutator algebra and single differential equation below
$$
[~ \nabla_+ \, , \, \nabla_+ ~ \} = i 2 \nabla_{\pp} ~~ , ~~ [~ \nabla_+ 
\, , \, \nabla_{\pp} ~ \}  = 0 ~~ , ~~ \nabla_+ {\S}^+ = \fracm 12 
{\cal R} \quad , 
$$
$$
[~ \nabla_{+} \, , \, \nabla_{\mm} ~\} = - i 2 {\S}^+ \cm  ~~ , ~~ 
[~ \nabla_{\pp} \,, \, \nabla_{\mm} ~ \}  = - ( ~ {\S}^+ \nabla_+ ~+~ 
{ \cal R } \cm ) ~ \quad . \eqno(16)
$$
The quantities ${\S}^+ $ and ${ \cal R }$ are field strength superfields 
and ${\cal M}$ denotes the generator of the 2D Lorentz group defined to 
act according to the rules; $[ {\cal M} , \psi_+ ] = \frac 12 \psi_+$, 
$[ {\cal M} , \psi_- ] = - \frac 12 \psi_-$, $[ {\cal M} , {\rm e}_{\pp} 
] = {\rm e}_{\pp}$ and $[ {\cal M} , {\rm e}_{\mm} ] = - {\rm e}_{\mm}$.  
On defining ${\S}^+ \slsh$ as the limit of $ {\S}^+ $ as the 
Grassmann coordinate is taken to zero and similarly for ${\cal R } \slsh$, 
we find 
$$
{\S}^+ \slsh ~=~ - {\psi}_{\pp, ~ \mm} {}^+ \,=\, - [~ {\rm e}_{\pp}
{\psi}_{\mm}{}^+ -  {\rm e}_{\mm}  {\psi}_{\pp}{}^+ - c_{\pp,~\mm}{}^{\pp}
{\psi}_{\pp}{}^+ - c_{\pp,~\mm}{}^{\mm} {\psi}_{\mm}{}^+ ~] ~ , \eqno(17)
$$
$$
{r}_{\pp,~\mm}( \o )  ~=~ - [~ {\rm e}_{\pp} {\o}_{\mm} -  {\rm e}_{\mm}
{\o}_{\pp} - c_{\pp,~\mm}{}^{\pp} {\o}_{\pp} - c_{\pp,~\mm}{}^{\mm}
{\o}_{\mm} ~] ~ ,  
$$
$$
{\nabla}_+ {\S}^+ \slsh  ~=~ - \fracm 12 [ ~ {r}_{\pp,~\mm}( \o ) + i 2 {
\psi}_{\pp} {}^+ { \psi}_{\pp, ~\mm} {}^+ ~ ] \quad , \eqno(18)
$$
$$
{\Hat \nabla}_{\pp}  \equiv {\rm e}_{\pp} + {\o}_{\pp} {\cal M} \qquad , 
\qquad {\Hat \nabla}_{\mm} \equiv {\rm e}_{\mm} + {\o}_{\mm} {\cal M} 
\qquad , \qquad  {\o}_{\pp} \,=\, c_{\pp,~\mm}{}^{\mm} \quad , 
$$
$$ 
~~{\o}_{\mm} \,=\, c_{\pp,~\mm}{}^{\pp} + i 2 {\psi}_{\pp}{}^+{\psi
}_{\mm}{}^+ ~~, ~~ {\rm e}_{a} \equiv {\rm e}_a {}^m {\pa}_m ~~ , ~~ 
[ {\rm e}_a , {\rm e}_b ] = c_{a, b} {}^c {\rm e}_c \quad . \eqno(19)
$$
But $\cal R$ is the vector-vector component of the super 2-form $R_{AB}$.  
Thus, we take the first equality in (17) and use it to replace the
${ \psi}_{\pp, ~\mm} {}^+ ~$ term on the last line of (18) to find,
$$
- \fracm 12 ~ {r}_{\pp,~\mm}( \o ) ~=~ \Big[ ~ \Big( ~ {\nabla}_+
~-~ i {\psi}_{\pp} {}^+ \Big)  {\S}^+ \, | ~\Big] ~~~. \eqno(20)
$$
This is a special case of (11) and following the general discussion
we enforce the E.C. by defining
$$
\Tilde \D ~\equiv~ - \fracm 12 ~ \int d^2 \s ~{\rm e}^{-1} \, 
{r}_{\pp,~\mm}( \o (e, \psi)) 
 ~~~, \eqno(21) 
$$
$$ 
{\Hat \D} ~\equiv~  ~ \int d^2 \s ~{\rm e}^{-1} \, \Big[ ~
\Big( ~ {\nabla}_+
~-~ i {\psi}_{\pp} {}^+ \Big)  {\S}^+ \, | ~\Big] ~~~. \eqno(22)
$$
and according to the E.I.T. and E.C. it must also be the case that
$$ \eqalign{
\int d^2 \s \, d \z^- ~{\rm E}^{-1} \, {\cal L}_- &\equiv~
~ \int d^2 \s ~{\rm e}^{-1} \, \Big[ ~ {\cal D}_+ \, 
{\cal L}_- \, | ~\Big] \cr
&= ~
\int d^2 \s ~{\rm e}^{-1} \, \Big[ ~ \Big( ~ {\nabla}_+
~-~ i {\psi}_{\pp} {}^+ \Big)  {\cal L}_- \, | ~\Big] 
~~~.} \eqno(23)
$$
exactly as stated in the first work of reference \cite{bgm}.  

Now the expression for ${\Tilde \D}$ in (21) allows us to calculate
the form of the terms in ${\Tilde \D}~=~\Delta~+~...$. This is
done by observing that
$$
{r}_{\pp,~\mm}( \o (e, \psi))  ~=~ {r}_{\pp,~\mm}( \o (e, 0))
~+~ i 2 \{ \nabla_{\pp}(e) [ \, \psi_{\pp}{}^+ \,  \psi_{\mm}{}^+ 
\, ] \, \} ~~~. \eqno(24)
$$ 
so that
$$
\Tilde \D ~\equiv~ - \fracm 12 ~ \int d^2 \s ~{\rm e}^{-1} \, 
{r}_{\pp,~\mm}( \o (e, 0)) ~-~ i \int d^2 \s ~{\rm e}^{-1} \,
\{ \, \pa_m [ \, {\rm e}_{\pp}{}^m (  \, \psi_{\pp}{}^+ \,  
\psi_{\mm}{}^+ \, ) \, ] \, \}
 ~~~. \eqno(25) 
$$
We see that the first term above is $\Delta~=~2\pi (g - 1)$
(where $g$ is the genus of the manifold), the usual
topological index on a 2-manifold,
$$
\D ~\equiv~ - \fracm 12 ~ \int d^2 \s ~{\rm e}^{-1} \, 
{r}_{\pp,~\mm}( \o (e, 0)) 
 ~~~, \eqno(26) 
$$
and the second term in (25) is what was indicated by $...$ in the E.C.  

Perhaps the reader was not impressed by the (1,0) example.  So let's 
repeat all of this in the more complicated case of 3D, N = 1 superspace.  
In 1979 \cite{bg} the superspace description of 3D, N = 1 
irreducible off-shell supergravity was first given
$$[ \nabla_{\a} ~,~ \nabla_{\b} \}  = i 2 (\g^c)_{\a \b} \left [ \nabla_c 
- R {\cal M}_c \right ] ~~~, ~~~~~
~~~~~~~~~~~~~~~~~~~~~~~~~~~~~~~~~~ $$
$$~[ \nabla_\a ~,~ \nabla_b \}  =  i (\g_b)_{\a}{}^{\d} [~ \fracm 12 R
\nabla_{\d} ~+~ ( \S_{\d} {}^d ~+~ i \fracm 23 (\g^d)_{\d}{}^{\e}
( \nabla_{\e} R )) {\cal M}_d ~] ~~~~~~~~ $$
$$~~~~~~~~~+~ ( \nabla_{\a} R ) {\cal M}_b  ~~~~~, ~~~~~~~~~~~~~~~~~~~~
~~~~~~~~~~~~~ $$
$$[ \nabla_a ~,~ \nabla_b \}  =   - \fracm 12
\e_{a b c } [~ \S^{\a c} + i \fracm 23 (\g^c)^{\a \b} (\nabla_{\b}
R) ~] \nabla_{\a} ~~~~~~~~~~~~~~~~~~~~~~~~~~ $$
$$
~~~~~~~~~~-~  \e_{a b c }[~  {\cal R}^{c d}  ~+~ \fracm 23 \eta^{c d} 
(\nabla^2 R ~-~ \fracm 32 R^2 )~] {\cal M}_d  ~~~, ~~~~~~
\eqno(26) $$
where ${\cal R}^{a b} - {\cal R}^{b a} = \eta_{a b} {\cal R}^{a b} = 
(\g_d)^{\a \b} \S_{\b}{}^d = 0$ and 
$$
 \nabla_{\a} \S_{\b}{}^c = i (\g_b)_{\a \b} {\cal R}^{b c} 
~-~ \fracm 23 [~ C_{\a \b} \eta^{c d} ~+~ i \fracm 12 (\g_b)_{\a \b} 
\e^{b c d} ~] \, (\nabla_d R)~~. 
\eqno(27) $$
In writing these results, their form was simplified by replacing
the usual Lorentz generator according to: $ {\cal M}_{ b c} 
\to \e_{b c }{}^a {\cal M}_a $, so that when acting on a spinor 
$\psi_{\a}$ or a vector $v_a$ we have
$$
[ \, {\cal M}_a ~,~ \psi_{\a} \, ] ~=~ i \fracm 12 (\g_a)_{\a}{}^{\b} 
\psi_{\b} ~~~,~~~ [ \, {\cal M}_a ~,~ v_b \, ] ~=~  \e_{a b }{}^{c} 
v_{c}   ~~~~. 
\eqno(28) $$

Now $R$ is a component of a super 2-form, but in 3D we need a super 3-form. 
Using the formalism of the 1981 work \cite{sjg1}, it is easy to show that 
an irreducible 3D, N = 1 closed super 3-form is described by ${\cal 
G}_{A B C}$ where
$$
{\cal G}_{\a \b \g}  ~=~  0 ~~~, ~~~
{\cal G}_{\a  \b c} ~=~ i 2 (\g_{ c })_{ \a \b} { {\cal G}} 
~~~, $$
$$
{\cal G}_{\a  b c} ~=~  i \e_{a b c} (\g^{ a })_{ \a}{}^{ \b} \, 
( \nabla_{\b} {\cal G} \,) ~~~, ~~~ {\cal G}_{a  b c}  ~=~ \e_{a b c} 
\, [\, \de^2 {\cal G} ~-~  R {\cal G} \, ] ~~~
~~~. \eqno(29)
$$
For this theory, the general result in (11) takes the form 
$$
{\cal G}_{a b c} | ~ = ~ {\Tilde g}_{a b c} ~+~ \e_{a b c} \Big[ 
\, i \psi_d {}^{\a} (\g^d )_{\a}{}^{ \b} ( \, \nabla_{\b} {\cal G}| 
\, ) ~-~ i  \e^{ d e f} \psi_d {}^{\a} (\g_e )_{\a \b} \psi_f {}^{\b}
(\, {\cal G} | \,) ~\Big]  ~~~~, \eqno(30)
$$
so that after substitution of the last result from (29) into the lhs of
(30) it follows that
$$
\fracm 16 \e^{a b c} {\Tilde g}_{a b c} ~= ~ ( {\cal D}^2 {\cal G} 
| \,)  ~~~,~~~  \eqno(31) 
$$
where the explicit form of the operator ${\cal D}^2$ is given by
$$
{\cal D}^2 ~ \equiv ~ \de^2  ~-~ i \psi_a {}^{\a} (\g^a )_{\a} 
{}^{ \b} \nabla_{\b} ~-~ R  ~+~ i  \e^{ a b c} \psi_a {}^{\a} (\g_b 
)_{\a \b} \psi_c {}^{\b}~~~.  \eqno(32)
$$
Therefore I am to define
$$
\Tilde \D ~\equiv~ \fracm 16 ~ \int d^3 x ~{\rm e}^{-1} \, 
 \e^{a b c} {\Tilde g}_{a b c}
 ~~~, ~~ \eqno(33) 
$$
$$ 
{\Hat \D} ~\equiv~  ~ \int d^3 x ~{\rm e}^{-1} \, \Big( ~
{\cal D}^2 {\cal G} \, | ~ \Big) ~~~, \eqno(34)
$$
and again using the E.I.T. and E.C. to define
$$
\int d^3 x ~d ^2 \q  \, {\rm E}^{-1} \, {\cal L} ~\equiv~ 
\int d^3 x ~{\rm e}^{-1} \, \Big[ ~ {\cal D}^2 {\cal L} 
\, | ~ \Big] ~~~. \eqno(35)
$$

Since ${\cal D}^{2}$ was derived via the E.C., we might want to check it 
on another choice of ${\cal L}$ such as ${\cal L} ~ =~ R$.  It is known for 
this choice that $S_{SG}~\propto~\int d^3 x ~ d^2 \q
{\rm E}^{-1} R$ is the correct answer.  After the usual projection 
techniques, I find
$$
\int d^3 x d^2 \q \, {\rm E}^{-1}\,  R = \int d^3 x {\rm e}^{-1} 
\left [ \, \, - \ha \e^{a b c} \left ( {\cal R}_{a b c} (\o) ~+~  
\psi_a {}_{\a} \Psi_{b c} {}^{\a} \right ) ~-~  B^2 ~ \right ]   ~~. 
\eqno(36)
$$
where the $ \Psi_{a b} {}^{\b}$ is the usual component level gravitino field
strength and the spin-connection is given by, 
$$
\o_a {}^b ~=~ \fracm 14   \e^{b c d} \left [ ~ C_{ c d a} ~-~ 2 C_{a c d}
~+~ i 4 \left (  \psi_c {}^{\a} (\g_a )_{\a \b} \psi_d {}^{\b} ~+~
 \psi_a {}^{\a} (\g_c )_{\a \b} \psi_d {}^{\b} \right ) ~  \right ]
~-~ \fracm 12 B \d_a {}^b ~~~.
\eqno(37) $$

Finally I have checked this same procedure using old minimal off-shell 
supergravity in 4D, N = 1 Wess-Zumino superspace to calculate the 
topological index ${\Hat \Delta}$ associated with the 4D, N = 1 super 
4-form multiplet described in the 1981 paper \cite{sjg1}. I find the result
$$
\eqalign{ 
{\Hat \D} ~=~ \int d^4 x ~ {\rm e}^{-1} \, \Big[ ~  - i \, (
\, {\cal D}^2 {\cal F} \, | )~+~ {\rm {h. \, c.}} ~\Big] ~~~,  }
\eqno(38) $$
here the operator ${\cal D}^2$ is defined by
$$
{\cal D}^2 ~\equiv ~ \nabla^2 ~+~ i  {\Bar \psi}{}^{\un a}{}_{\dot \a} 
\nabla_{\a} ~+~ 3 {\Bar R} ~+~  \fracm 12 C^{\a \b} {\Bar \psi}_{\un a }
{}^{( \dot \a} \, {\Bar \psi}_{\un b  }{}^{\dot \b )} 
~~~, \eqno(39) 
$$
and $\cal F$ is the lowest non-trivial super 4-form field strength
component
$$
F_{\a \, \b \, \un c \, \un d} ~=~  C_{\Dot \g \Dot \d} C_{\a (\g}
C_{\d ) \b} {\Bar {\cal F}} ~~~. \eqno(40)
$$
I note that the form of (38) suggests the formula
$$
\int d \m ~ {\cal L}_{Gen} ~=~ \fracm 12 \, \int d \m_c ~ {\cal L}_c 
~+~ {\rm {h.\, c.}} ~~~. \eqno(41)
$$
So that the E.C. and E.I.T. imply
$$
\int d \m_c ~ {\cal L}_c ~=~ \int d^4 x \, {\rm e}^{-1} ~ \Big[ \,
{\cal D}^2 {\cal L}_c \, | ~ \Big] ~~~,
\eqno(42)
$$
acting on a chiral superfield (such as ${\cal F}$).  More generally
$$
{\cal L}_{c}~=~({\Bar \nabla}^{2} ~+~ R) \, {\cal L}_{Gen} ~~~, 
\eqno(43)
$$
so that we may define
$$ \eqalign{
\int d \m ~{\cal L}_{Gen}  &=~ \fracm 12\Big\{ ~ \int d^4 x \, {\rm 
e}^{-1} ~ \Big[ \, {\cal D}^2 \, ~({\Bar \nabla}^{2} ~+~ R) \, {\cal 
L}_{Gen} \, | ~ \Big] ~+~ {\rm {h.c.}} ~\Big\} \cr
&\equiv~ \int d^4 x \, {\rm e}^{-1} ~ \Big[ \, {\cal D}^4 ~ {\cal 
L}_{Gen} \, | ~ \Big] ~~~, }
\eqno(44) $$
where the operator ${\cal D}^4$ is defined by
$$
{\cal D}^4 ~=~ \fracm 12 \, \Big[ ~ {\cal D}^2 \, ~({\Bar \nabla}^{2} 
~+~ R) ~+~ {\Bar {\cal D}}^2 \, ~({ \nabla}^{2} ~+~ {\Bar R}) ~\Big] 
~~~. \eqno(45) $$
This final result can be seen to coincide exactly with the result 
in our book `{\it {Superspace}}' where it was `derived by a handicraft' 
argument.

Thus, I see that there is excellent support for the E.I.T. and E.C. from a 
number of explicit cases.  I have also found numerous other examples.  
I am still checking even more examples in an attempt to understand if 
there are any limitations on this method.

I think the E.C. is a universal feature of all supergravity theories that 
has escaped our notice since the beginning of the era of using Wess-Zumino 
superspace!  I also have some evidence that the E.~C. plays an even more
important role than I presented here.  In some examples I know, there
occur topological obstructions to the imposition of the E.~C.  The most
interesting point about these obstructions is that they take the
form of the supergravity constraints themselves!  It is perhaps
not too optimistic to hope that at last we have begun to grasp the
`deep' reason why constraints must be imposed in supersymmetrical
theories. The answer seems to be to enforce the E.~C.

I further conjecture that the E.C. will ultimately be found 
to apply to even covariant string field theory!  The reasoning
goes as follows.  In a fully covariant and geometrical approach to
string and superstring field theory, one must be confronted with
calculating the integral $\int d^D {\bf X}(\s) \, d {\bf B}(\s)
\, d {\bf C}(\s)$ (here we consider the bosonic string for
the sake of simplicity). It ought to be possible to write an
equation like
$$
\int d^D {\bf X}(\s) \,\, d {\bf B}(\s) \,\, d {\bf C}(\s)
~=~ \int d^D {\bf X}(0) ~ {\cal D}^{(\infty)} ~~~, \eqno(46) $$
so that the string coordinate zero-modes define the manifold of an ordinary
appearing field theory.  The oscillator modes (of all types) define 
the ectoplasm of the string space.  Thus in a fully geometrical approach
to covariant string field theory I expect that there should exist an
operator ${\cal D}^{(\infty)}$ that appears in a `stringy'
E.I.T.

If this conjecture proves to be true, it provides an elegantly simple basis 
for understanding why even if we live in a universe described by fiber 
bundles, Kaluza-Klein spaces, Wess-Zumino superspace, strings,
superstrings, heterotic strings, branes, M-theory, F-theory etc., the 
topological triviality of all the extra ``coordinates'' may forbid their 
having direct physical consequences (at least in the absence of strong
coupling given the current views of strong/weak duality).  The Ethereal 
Conjecture, properly interpreted, may be a physical principle.

\noindent {\bf{Acknowledgment} }
\indent \newline
${~~~~}$I wish to thank the organizers of the International Seminar on
``Supersymmetries and Quantum Symmetries'' for their kind invitation
to deliver this presentation at this memorial meeting to honor
Dr. Ogievetsky. Additional thanks go to T.~H\' ubsch and M.~Luty
for their assistance in the preparation of this manuscript.

\newpage

\end{document}